\documentclass[prb,amsfonts,showpacs,preprint]{revtex4}
\usepackage{graphics,bm,natbib}
\begin{document}
\widetext
\title{Screening of Coulomb interactions in transition metals}
\author{I. V. Solovyev}
\email[Electronic address: ]{igor@issp.u-tokyo.ac.jp}
\author{M. Imada}
\affiliation{
Institute for Solid State Physics, University of Tokyo,\\
Kashiwanoha 5-1-5, Kashiwa, Chiba 277-8531, Japan, and\\
PRESTO, Japan Science and Technology Agency
}
\date{\today}

\widetext
\begin{abstract}
We discuss different methods of calculation of the screened Coulomb interaction $U$
in transition metals and compare the
so-called constraint local-density approximation (LDA) with the GW approach.
We clarify that they offer complementary methods of treating the
screening and, therefore, should serve for
different purposes. In the \textit{ab initio} GW method,
the renormalization
of \text{bare} on-site Coulomb interactions between $3d$ electrons (being of the
order of 20-30 eV)
occurs mainly through the screening by the same $3d$ electrons, treated
in the
random phase approximation (RPA).
The basic difference of the constraint-LDA method from the GW method
is that it deals with the
neutral processes, where the Coulomb interactions are additionally screened by
the ``excited'' electron, since it \textit{continues to stay in the system}.
This is the main channel of screening by the itinerant ($4sp$) electrons, which
is especially strong in the case of transition metals and missing in the
GW approach, although the details of this screening may be affected by
additional approximations, which typically supplement these two methods.
The major drawback
of the conventional constraint-LDA method is that it does not
allow to treat the energy-dependence of $U$, while the full GW calculations require
heavy computations. We propose
a promising approximation based on the combination
of these
two methods.
First, we take into account the screening of Coulomb interactions
in the $3d$-electron-line bands located near the Fermi level
by the states from the subspace being orthogonal to these bands, using
the constraint-LDA methods. The obtained interactions
are further renormalized within the bands near the Fermi level
in RPA. This allows
the energy-dependent screening
by electrons near the Fermi level including the same $3d$ electrons.
\end{abstract}



\maketitle


\section{\label{sec:intr}Introduction}

  The description of electronic structure and properties of strongly correlated
systems presents a great challenge for
\textit{ab initio} electronic structure
calculations.
The main complexity of the problem is related with the fact that such electronic systems
typically bear both localized and itinerant character, where most
of conventional methods do not apply.
A canonical example is the local-[spin]-density approximation (L[S]DA) in the
density-functional theory (DFT).\cite{DFT}

  The DFT, which is a ground-state theory, is based on the minimization of the total
energy functional $E[\rho]$ with respect to the electron density $\rho$.
In the Kohn-Sham (KS) scheme,
which is typically employed for practical calculations,
this procedure is formulated as the self-consistent solution of
single-particle KS equations
\begin{equation}
\left( -\nabla^2 + V_{\rm KS}[\rho] \right) \psi_i[\rho] = \varepsilon_i \psi_i[\rho],
\label{eqn:KS}
\end{equation}
which are combined
with the equation for the electron density:
\begin{equation}
\rho = \sum_i f_i |\psi_i|^2,
\label{eqn:rho}
\end{equation}
defined in terms of
eigenfunctions ($\psi_i$), eigenvalues ($\varepsilon_i$), and the
occupation numbers ($f_i$) of KS quasiparticles.

  The LSDA provides an explicit expression for $V_{\rm KS}[\rho]$. However, it
is based on the homogeneous
electron gas model, and strictly speaking
applicable only for itinerant electron compounds.

  The recent progress, which
gave rise to such directions as LDA$+$ Hubbard $U$ (Refs.~\onlinecite{AZA,PRB94,LDAUreview}) and
LDA+DMFT (dynamical mean-field theory) (Refs.~\onlinecite{LSDADMFT,LichtPRL01}), is based on the
idea of partitioning of electronic states.
It implies the validity of the following postulates: \\
(1) All solutions of KS equations (\ref{eqn:KS}) in LDA can be divided
(by introducing proper projection-operators) into two subgroups:
$i$$\in$$I$, for which LSDA works reasonably well, and $i$$\in$$L$, for which
LSDA encounters serious difficulties and needs to be improved (a typical example
is the $3d$ states in transition-metal oxides and some transition metals). \\
(2) Two orthogonal subspaces, $I$ and $L$,
are ``flexible''
in the sense that they
can be defined for a wider
class of electron densities, which can be
different from the ground-state density in LDA.
This allows to ``improve'' LDA by adding a proper correction $\Delta\hat{\Sigma}$
(generally, an $\omega$-dependent self-energy)
to the KS equations, which acts solely in the $L$-subspace but may also affect the
$I$-states through the change of $\rho$ associated with this $\Delta\hat{\Sigma}$.
Thus, in the KS equations, the $L$- and $I$-states remain decoupled even after including
$\Delta\hat{\Sigma}$:
$\langle \psi_{i \in I}[\rho] |(-\nabla^2$$+$$V_{\rm KS}[\rho]$$+$$
\Delta\hat{\Sigma})| \psi_{i \in L}[\rho] \rangle$$=$$0$.
For many applications, the $L$-states are atomic or Wannier-type orbitals.
In this case, the solution of the problem in the $L$-space becomes equivalent
to the solution of a multi-orbital Hubbard-type model, and the formulation of the
LDA$+$$U$ approach is basically a mapping of the electronic structure
in LDA onto this Hubbard model. In the following, by referring to the LDA$+$$U$ we will
mean not only the static version of this method, originally proposed in Ref.~\onlinecite{AZA},
but also its recent extensions designed to treat dynamics of
correlated electrons and employing the same idea of
partitioning of the electronic states.\cite{LSDADMFT,LichtPRL01}\\
(3) All physical interactions, which contribute to $\Delta\hat{\Sigma}$, can be formally
derived from LDA by introducing certain constraining fields $\{ \delta \hat{V}_{\rm ext} \}$
in the subspace of $L$-states of
the KS equations (i.e., in a way similar to $\Delta\hat{\Sigma}$). The purpose of
including these
$\{ \delta \hat{V}_{\rm ext} \}$ is to simulate
the change of the electron density, $\delta \rho$, and then to extract parameters of electronic
interactions from the total energy difference $E[\rho$$+$$\delta \rho]$$-$$E[\rho]$,
by doing a mapping onto the Hubbard model.
The total energy difference is typically evaluated in LDA,\cite{JonesGunnarsson}
and the method itself is called the constraint-LDA
(CLDA).\cite{UfromconstraintLSDA,Gunnarsson,AnisimovGunnarsson,PRB94.2}

  However, despite a more than decade of rather successful history, the central
question of LDA$+$$U$ is not completely solved and continues to be the subject
of various disputes and controversies.\cite{NormanBrooks,PRB96,Pickett,Springer,Kotani,Ferdi}
This question is how to define the parameter of the effective Coulomb
interaction $U$.

  To begin with,
the Coulomb $U$
is \textit{not} uniquely defined quantity, as it strongly depends on the property
for the description of which we want to correct our LDA scheme.
One possible strategy is the excited-state properties, associated with
the complete removal of an electron from (or the addition of the new electron to)
the system, i.e. the processes which are described by Koopman's theorem in Hartree-Fock
calculations and which are corrected in the GW method by taking into account
the relaxation of the wavefunctions onto the created electron hole
(or a new electron).\cite{Hedin,FerdiGunnarsson}
However the goal which is typically pursued in LDA$+$$U$
is somewhat different. Namely, one would always like to stay as close as
it is possible to the description of the ground-state properties. The necessary precondition
for this, which should be taken into account in the definition of the Coulomb $U$
and all other interactions which may contribute to $\Delta \hat{\Sigma}$ is the
conservation of the total number of particles. In principle, similar strategy can
be applied for the analysis of neutral excitations (e.g., by considering the
$\omega$-dependence of $\Delta \hat{\Sigma}$), for which the total number of electrons
is conserved.\cite{LichtPRL01}
The basic difference between these two processes is that the ``excited'' electron in the
second case continues to stay in the system and may additionally screen the Coulomb $U$.
This screening may also affect the relaxation effects.\cite{OnidaReiningRubio}

  The purpose of this paper is to clarify several questions related with the
definition of the Coulomb interaction $U$ in transition metals. We will discuss both
the
momentum (${\bf q}$) and energy ($\omega$) dependence of $U$, corresponding to the response
of the Coulomb potential
onto the site (${\bf R}$)
and time ($t$) dependent perturbation $\delta \hat{V}_{\rm ext}$, and present a comparative
analysis of the existing methods of calculations of this interaction, like CLDA and GW.
We will argue that, despite a common believe,
the GW method does not take into account
the major effect of screening
of the effective Coulomb interaction $U$ between the $3d$ electrons
by the (itinerant) $4sp$ electrons, which may also
contribute to the ${\bf q}$-dependence of $U$.
This channel of screening is
included in CLDA,
although under an additional approximation separating the $3d$- and $4sp$-states,
while in the GW approach, its absence can be compensated
by an appropriate choice of the pseudo-Wannier orbitals, simulating the basis of
$L$-states.
On the other hand, CLDA is a static approach, which does not
take into account the $\omega$-dependence of $U$.\cite{TDDFT}
We will consider mainly the ferromagnetic (FM) fcc Ni,
although similar arguments can be applied for other metallic compounds.
We start with the basic definition of $U$ for the systems with the conserving
number of particles, which was originally introduced by Herring,\cite{Herring}
and then discuss the connection of this definition with the parameters which comes out
from CLDA and GW calculations.

\section{\label{sec:Herring}Herring's definition and CLDA}

  According to Herring,\cite{Herring} the Coulomb $U$ is nothing but the
energy cost for moving a $L$-electron between two atoms, located at ${\bf R}$
and ${\bf R}'$, and initially populated by $n_{L{\bf R}}$$=$$n_{L{\bf R}'}$$\equiv$$n_L$
electrons:
\begin{equation}
U_{{\bf RR}'} = E [ n_{L{\bf R}}+1,n_{L{\bf R}'}-1 ] -
E [ n_{L{\bf R}},n_{L{\bf R}'}  ].
\label{eqn:HerringU1}
\end{equation}
In DFT, $U_{{\bf RR}'}$ can be expressed in terms of the KS eigenvalues,
$\varepsilon_{L{\bf R}}$$=$$\partial E/\partial n_{L{\bf R}}$, using Slater's
transition state arguments:\cite{PRB94.2}
\begin{equation}
U_{{\bf RR}'} =
\varepsilon_{L{\bf R}} [ n_{L{\bf R}}+\frac{1}{2},n_{L{\bf R}'}-\frac{1}{2} ] -
\varepsilon_{L{\bf R}} [ n_{L{\bf R}}-\frac{1}{2},n_{L{\bf R}'}+\frac{1}{2} ].
\label{eqn:HerringU1TS}
\end{equation}
The final definition
\begin{equation}
U_{{\bf RR}'} = \left. \frac{\partial \varepsilon_{L{\bf R}}}{\partial n_{L{\bf R}}}
\right|_{n_{L{\bf R}}+n_{L{\bf R}'}=const},
\label{eqn:HerringU1TSD}
\end{equation}
which is typically used in CLDA calculations, is obtained after replacing the finite
difference between two KS eigenvalues in Eq.~(\ref{eqn:HerringU1TS})
by their derivative. The derivative
depends on the path in the
sublattice of occupation numbers
along which it is calculated
(e.g., $n_{L{\bf R}}$$+$$n_{L{\bf R}'}$$=$$const$).
This dependence has a clear physical meaning and originates from
the distance-dependence of
intersite Coulomb interactions, which contribute to the screening of $U_{{\bf RR}'}$.
In the reciprocal (Fourier) space, this distance-dependence
gives rise to the ${\bf q}$-dependence of $U$.

  Owing to the existence of the second subsystem, $I$, the reaction (\ref{eqn:HerringU1})
may compete with another one
\begin{equation}
U = E [ n_{L{\bf R}}+1,n_{I{\bf R}}-1,n_{L{\bf R}'}-1,n_{I{\bf R}'}+1  ] -
E [ n_{L{\bf R}},n_{I{\bf R}},n_{L{\bf R}'},n_{I{\bf R}'}  ],
\label{eqn:HerringU2}
\end{equation}
corresponding to independent ''charge transfer'' excitations at the sites
${\bf R}$ and ${\bf R}'$.\cite{ZSA}
It can be also presented in the form (\ref{eqn:HerringU1TSD}),
but with the different constraint imposed on the numbers of
$L$- and $I$-electrons: $n_{L{\bf R}}$$+$$n_{I{\bf R}}$$=$$const$.
Generally, the definitions (\ref{eqn:HerringU1}) and (\ref{eqn:HerringU2})
will yield two different interaction parameters. Since
in the charge-transfer scenario
any change
of $n_{L{\bf R}}$ is totally screened by the change of $n_{I{\bf R}}$ located at the same site,
the interaction (\ref{eqn:HerringU2}) does not depend on ${\bf R}$.

  In reality, both processes coexist and the proper interaction parameter is given by
the following equation
$$
U_{{\bf RR}'} = E [ n_{L{\bf R}}+1,n_{I{\bf R}}-\delta,n_{L{\bf R}'}-1,n_{I{\bf R}'}+\delta  ] -
E [ n_{L{\bf R}},n_{I{\bf R}},n_{L{\bf R}'},n_{I{\bf R}'}  ],
$$
where the amount of charge $\delta$ redistributed between two subsystems is determined variationally
to minimize $U_{{\bf RR}'}$.
In
the CLDA scheme, it is convenient to work in the reciprocal (Fourier) space and calculate
$U_{\bf q}$
as the response to the ${\bf q}$-dependent constraining field
\begin{equation}
\delta \hat{V}_{\rm ext}({\bf q},{\bf R})=V_L \cos {\bf q} \cdot {\bf R},
\label{eqn:dVext}
\end{equation}
acting in
the subspace of $L$-states under the general condition of
conservation of the total number of particles.
The results of these calculations will strongly depend on how well
$L$-electrons are screened by the $I$-ones. In the case of perfect (100\%) screening,
the reaction (\ref{eqn:HerringU2}) will dominate, and the parameter $U$ will not depend
on ${\bf q}$. If the screening is not perfect (e.g., the change of the number of $3d$
electrons in the transition metals is screened to only about 50\% by the $4sp$ electrons
at the same atom -- Ref.~\onlinecite{AnisimovGunnarsson}),
it is reasonable to expect strong ${\bf q}$-dependence of the effective $U$,
because two different channels of screening, given by Eqs.~(\ref{eqn:HerringU1}) and
(\ref{eqn:HerringU2}), will work in a different way for different ${\bf q}$'s.
Since the excess (or deficiency) of $L$-electrons caused by a uniform shift of
the external potential $\delta \hat{V}_{\rm ext}$ can be only compensated from the system of
$I$-electrons, the ''charge transfer'' mechanism (\ref{eqn:HerringU2}) will always
dominate for small ${\bf q}$.
The mechanism (\ref{eqn:HerringU1}) becomes increasingly important near
the Brillouin zone (BZ) boundary, and will generally compete with the ''charge transfer'' excitations
(\ref{eqn:HerringU2}),
depending on the distribution of the $I$-electron density.~\cite{AnisimovGunnarsson}

\section{\label{sec:GWg}The GW method}

  It was recently suggested by several authors
(e.g., in Refs. \onlinecite{LDAUreview,Springer,Kotani,Ferdi},
and \onlinecite{Biermann})
that the Coulomb $U$ in the LDA$+$$U$ approach can be replaced
by the screened Coulomb interaction $W$ taken from the \textit{ab initio} GW method.
The latter
is calculated
in the random phase approximation (RPA):\cite{Springer,Kotani,Ferdi}
\begin{equation}
\hat{W}(\omega) = \left[1 - \hat{u} \hat{P}(\omega)\right]^{-1} \hat{u}.
\label{eqn:Dyson}
\end{equation}
We adopt the orthogonal atomic-like basis of linear-muffin-tin orbitals (LMTO)
$\{ \chi_\alpha \}$,\cite{LMTO} which specifies all matrix notations in
Eq.~(\ref{eqn:Dyson}).
For example,
the matrix of bare Coulomb interactions
$e^2/|{\bf r}$$-$${\bf r}'|$ has the form
$\langle \alpha \beta | \hat{u} | \gamma \delta \rangle$$=$$
e^2 \int d{\bf r} \int d{\bf r}' \chi_\alpha^*({\bf r}) \chi_\beta^*({\bf r}')
|{\bf r}$$-$${\bf r}'|^{-1} \chi_\gamma({\bf r}) \chi_\delta({\bf r}')$,
and all other matrices are defined in a similar way.
The diagonal part of $\hat{u}$ for the $3d$ states is totally specified by
three radial Slater's integrals: $F^0$, $F^2$, and $F^4$.
In the following we will identify $F^0$ with the parameter of bare Coulomb
interaction, which has the same meaning as the Coulomb $U$ after taking
into account all screening effects.
$F^2$ and $F^4$ describe non-spherical interactions, responsible for Hund's rule.

  The first advantage of RPA is that it allows to handle the
$\omega$-dependence of $\hat{W}$, which comes from the $\omega$-dependence of the
polarization matrix $\hat{P}$. The most common approximation for $\hat{P}$,
which
is feasible for \textit{ab initio} GW calculations, is that of
non-interacting quasiparticles:\cite{Hedin,FerdiGunnarsson}
\begin{equation}
P_{\rm GW}({\bf r},{\bf r}',\omega) = \sum_{ij}
\frac{(f_i-f_j)\psi_i ({\bf r}) \psi^*_i ({\bf r}')
\psi^*_j ({\bf r}) \psi_j ({\bf r}')}
{\omega - \varepsilon_j + \varepsilon_i + i\delta (f_i-f_j)},
\label{eqn:polarization}
\end{equation}
which is typically evaluated starting with the electronic structure in LSDA
(the spin indices are already included in the definition of $i$ and $j$).
Generally speaking, the use of $\hat{P}_{\rm GW}$
is an additional approximation, which
yields a new interaction $\hat{W}_{\rm GW}$.
At this stage, it is not clear whether it has the same meaning as the effective $U$
derived from CLDA and whether
Eq.~(\ref{eqn:polarization}) includes all necessary channels of screening.
It may also
include some other effects, which should be excluded from the final definition of $U$,
in order to avoid the double-counting.
One is the
self-screening arising from local (on-site) interactions between the
localized electrons.
These interactions are not accurately treated in RPA.\cite{Liebsch}
Therefore, the basic idea is to exclude these effects
from the definition of $\hat{W}_{\rm GW}$ and to resort this part to the interaction term of
the Hubbard model.\cite{Biermann}
In this respect, the second important
property of RPA is that it allows to easily partition different
contribution to $\hat{P}$ and $\hat{W}$. If $\hat{P}$$=$$\hat{P}_1$$+$$\hat{P}_2$ and
$\hat{W}_1$ is the solution of Eq.~(\ref{eqn:Dyson}) for
$\hat{P}$$=$$\hat{P}_1$, the total $\hat{W}$ can be obtained from
the same equation after substitution
$\hat{P}$$\rightarrow$$\hat{P}_2$ and $\hat{u}$$\rightarrow$$\hat{W}_1$ in Eq.~(\ref{eqn:Dyson}).
For example, if $\hat{P}_2$$=$$\hat{P}_{LL}$
is the part of $\hat{P}_{\rm GW}$, which includes all possible transitions between
the localized
states, and $\hat{P}_1$$=$$\hat{P}_r$ is the rest of the polarization,
the matrix
$\hat{W}_r$ corresponding to $\hat{P}_r$, can be used as the interaction part
of the Hubbard model.\cite{Kotani,Ferdi}

\subsection{\label{sec:GWNi}The GW story for fcc Ni}

  The ferromagnetic fcc Ni is the most notorious example where LSDA encounters
serious difficulties, especially for description of
spectroscopic properties. There are three major problems:\cite{FerdiGunnarsson}
(i) the bandwidth is too large (overestimated by $\sim$30\%);
(ii) the exchange splitting is too large (overestimated by $\sim$50\%);
(ii) the absence of the 6 eV satellite.
The \textit{ab initio} GW approach corrects only the bandwidth (although with
certain tendency to overcorrect), whereas the other two problems remain even in
GW.\cite{FerdiGunnarsson,YamasakiFujiwara}
Therefore, before doing any extensions on the basis of GW method, it is
very important to have a clear idea about its limitations. In this section we
would like to clarify several confusing statements about screening of
$W$ in GW. We argue that
the main results of the \textit{ab initio} GW method
can be explained, even quantitatively,
by retaining, instead of the full matrix $\hat{u}$ in Eq.~(\ref{eqn:Dyson}),
only the site-diagonal block
$\hat{u}_{LL}$ of bare Coulomb interactions between $3d$ electrons, in the
atomic-like LMTO basis set.
An intuitive reason behind this observation is the form of polarization matrix
(\ref{eqn:polarization}), which can interact only with exchange matrix elements.
The latter are small unless they
are calculated between orbitals of the same type, corresponding to the
self-interaction.
The values of
radial Slater's integrals calculated in the basis of atomic $3d$ orbitals are
$F^0$$=$$24.9$, $F^2$$=$$11.1$, and $F^4$$=$$6.8$ eV, respectively.
All other interactions are considerably smaller. Hence, it seems to be reasonable to
adopt the limit $\hat{u}_{LL}$$\rightarrow$$\infty$, which automatically picks up
in Eq.~(\ref{eqn:Dyson}) only those matrix elements which are projected
onto the atomic $3d$ orbitals, in the LMTO representation.
In this sense
the \textit{ab initio} GW method for transition metals can be regarded
as the RPA solution of the Hubbard model with the \textit{bare} on-site interactions between
$3d$ electrons \textit{defined in the basis of LMTO orbitals}.
In the GW method, these interactions are practically not screened
by outer electrons.
Note, however, that the LMTO basis in the transition metals is generally different from
the Wannier basis, which should be used for the construction
the Hubbard Hamiltonian. As it will become clear in Sec.~\ref{sec:OQ},
the Wannier representation has several additional
features, which may modify conclusions of this section to a certain extent.

\begin{figure}[h!]
\begin{center}
\resizebox{12.0cm}{!}{\includegraphics{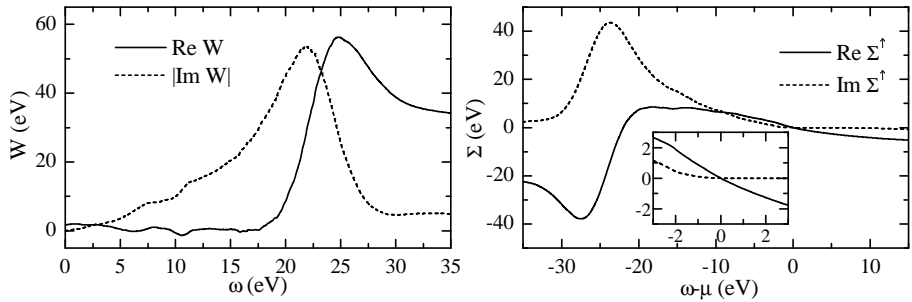}}
\end{center}
\caption{\label{fig.WandSb}
Characteristic behavior of
site-diagonal element of the screened Coulomb interaction
$W$$=$$\langle {\rm xy}~{\rm xy}|\hat{W}_{{\bf R}=0}| {\rm xy}~{\rm xy} \rangle$
and the matrix element of the
self-energy $\Sigma$$=$$\langle {\rm xy} | \hat{\Sigma}^\uparrow_{{\bf q}=0} | {\rm xy} \rangle$
between xy orbitals of the $t_{2g}$ manifold
in the $\Gamma$-point of Brillouin zone obtained in the GW approach
with the bare Coulomb interactions
between $3d$ electrons in the atomic-like LMTO basis set. Inset shows amplified
$\Sigma(\omega)$ near $\omega$$=$$\mu$.
Matrix elements between $e_g$ orbitals show a
similar behavior.}
\end{figure}

  Results of these model GW calculations are shown in Fig.~\ref{fig.WandSb}.
In this case, the energy scale is controlled by
the bare interaction $F^0$,
which predetermines the asymptotic behavior
${\rm Re}W(\infty)$ (with $W$ denoting the diagonal
matrix element of $\hat{W}$) and the position
of the ''plasmon peak'' of ${\rm Im} W(\omega)$ at $\sim$22 eV , which is
related with the sharp increase of ${\rm Re}W(\omega)$ at around
25 eV via the Kramers-Kronig transformation.
At small $\omega$, the behavior of $\hat{W}(\omega)$ is well
consistent with the
strong coupling regime $F^0$$\rightarrow$$\infty$: namely,
$\hat{W}(\omega)$$\sim$$-$$\hat{P}^{-1}(\omega)$,
which is small ($\sim$1.8 eV at $\omega$$=$$0$) and \textit{does
not depend on }$F^0$ (though it may
depend on $F^2$ and $F^4$).
All these features are in a good semi-quantitative agreement
with results of GW calculations.\cite{FerdiGunnarsson,Springer,Kotani,Ferdi}

\begin{figure}[h!]
\begin{center}
\resizebox{10.0cm}{!}{\includegraphics{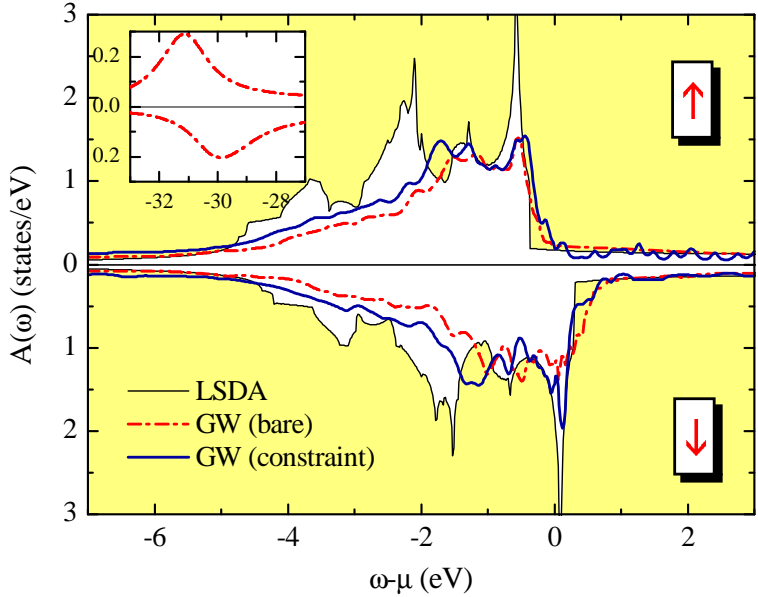}}
\end{center}
\caption{\label{fig.DOS}
The spectral function
$A(\omega)$$=$$-$$\frac{1}{\pi} {\rm Im} {\rm Tr} \hat{G}(\omega) {\rm sgn}(\omega$$-$$\mu)$
for fcc Ni in LSDA and two GW schemes with bare electronic interactions
and parameters extracted from constraint-LDA.
The inset shows the satellite structure in $A(\omega)$
at the $\Gamma$-point of Brillouin zone
in the bare-GW approach.}
\end{figure}

  The self-energy in GW is given by the convolution of $\hat{W}$ with the
one-particle Green function $\hat{G}$:
\begin{equation}
\hat{\Sigma}(\omega) = \frac{i}{2\pi} \int d\omega'
\hat{G}(\omega + \omega') \hat{W}(\omega').
\label{eqn:SigmaGW}
\end{equation}
Therefore, the $\omega$-dependence of $\hat{\Sigma}$
should incorporate the main features of $\hat{W}(\omega')$.
Indeed, the low-energy part of $\hat{\Sigma}$ (close to the Fermi energy or the chemical
potential $\mu$)
is mainly
controlled by ${\rm Im} \hat{W}$.
Since the main poles of
${\rm Im} \hat{W}$ and
${\rm Im} \hat{G}$
are well separated on the $\omega$-axis
(the $\omega$-range of ${\rm Im} \hat{G}$ is limited by the
$3d$ bandwidth, $\sim$4.5 eV in LSDA for fcc Ni, whereas the ''plasmon peak''
of ${\rm Im} W$ is located only at $\sim$22 eV), one has the following relation:
\begin{equation}
\left. \partial \Sigma /\partial \omega \right|_{\omega=\mu}
\approx \frac{1}{\pi} \int_0^\infty d\omega {\rm Im} W(\omega)/\omega^2.
\label{eqn:dSigmamu}
\end{equation}
This yields the renormalization
factor $Z$$=$$[1$$-$$\left.  \partial \Sigma /
\partial \omega \right|_{\omega=\mu}]^{-1}$$\sim$$0.5$, which readily explains
the reduction of the $3d$ bandwidth as well as of the
intensity of the valence spectrum in \textit{ab initio} GW calculations
(Fig.~\ref{fig.DOS}).\cite{FerdiGunnarsson,YamasakiFujiwara}

  Away from the Fermi energy (i.e., for energies $|\omega|$ which are
much larger than the $3d$ bandwidth), one has another relation
${\rm Re} \Sigma(\omega)$$\sim$$-$${\rm Re} W(\omega)$, which readily explains
the existence of the
deep minimum of ${\rm Re} \Sigma(\omega)$ near $-$$30$ eV as well as
large transfer of the spectral weight into
this region (shown in the inset of Fig.~\ref{fig.DOS}).
Therefore, it is not quite right to say that the satellite structure is
missing in the \textit{ab initio}
GW approach. It may exist, but only in the wrong region of $\omega$.

  Thus, even besides RPA, the major problem of the GW description for the transition
metals is the \textit{wrong energy scale, which is controlled by the bare on-site Coulomb
interaction} $F^0$ ($\sim$$20$-$30$ eV) between the $3d$ electrons.
In summarizing this section we would like to stress again the following points:\\
(1)
The major channel of screening of Coulomb interaction in the GW method for the
transition metals originates from
the $3d$$\rightarrow$$3d$ transitions in the polarization function calculated in the
atomic-like LMTO basis set.
The screening by the $4sp$-electrons is practically absent;\\
(2)
At small $\omega$,
the
deficiency of the $3d$-$4sp$ screening is masked
by the strong-coupling regime realized in RPA equations for screened Coulomb interaction,
which explains a
small value of $W(0)$ obtained in the GW calculations; \\
(3)
The main $\omega$-dependence of $\hat{\Sigma}$ and $\hat{W}$ in GW
also comes from the $3d$$\rightarrow$$3d$ transitions.

  Different conclusions obtained in Refs.~\onlinecite{Kotani,Ferdi}
are related with the use of different partitioning
into what is called the ``$3d$''
and ``non-$3d$'' (pseudo-) Wannier orbitals.\cite{comment3}
In the light of analysis presented in this section, the strong $\omega$-dependent
screening by the ``non-$3d$'' Wannier states obtained in Refs.~\onlinecite{Kotani,Ferdi}
means that in reality these states had a substantial weight of ``$3d$'' character
of the LMTO basis,
which mainly contributed to the screening.
We will return to this problem in Sec.~\ref{sec:OQ}.

\begin{figure}[h!]
\begin{center}
\resizebox{10.0cm}{!}{\includegraphics{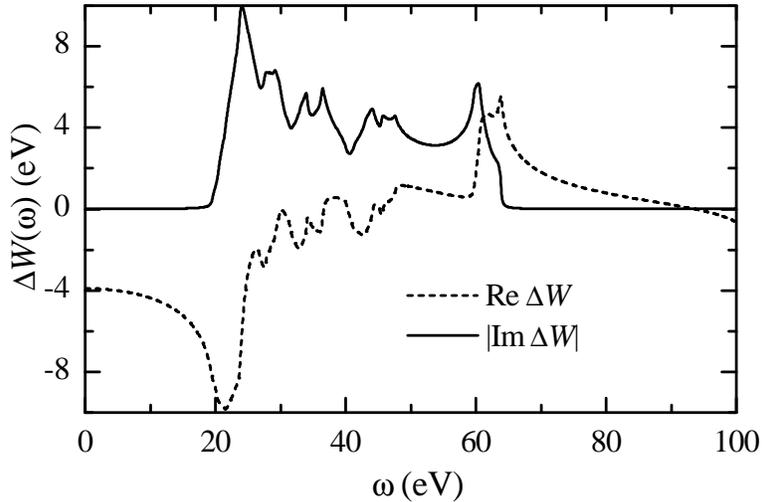}}
\end{center}
\caption{\label{fig.relaxation}
The $\omega$-dependence of
on-site Coulomb interaction associated with the relaxation of the
3d wavefunctions in the region of the $3d$$\rightarrow$$4d$ transitions.
The $3d$$\rightarrow$$5d$ transitions have been also taken into account.
They contribute to the region above 100 eV, which is not shown here.}
\end{figure}

  The next important
interaction, which contribute to the screening of $F^0$ in GW is due to
transitions between states with the same angular momentum: i.e.,
$3d$$\rightarrow$$nd$ ($n$$=$ $4$, $5$, ...)
(see also comments in Sec.~\ref{sec:conventions}).
In the lowest order (non-self-consistent RPA), these contributions can be
estimated as
\begin{equation}
\Delta W(\omega) \approx \langle 3d 3d | \hat{u} | 3d 4d \rangle_{\rm av}^2
P_{\rm GW}(\omega,3d \rightarrow 4d) + ({\rm higher}~n),
\label{eqn:simpleDyson}
\end{equation}
where $\langle 3d 3d | \hat{u} | 3d 4d \rangle_{\rm av}$$\simeq$$6.1$ eV
is the spherical part of the exchange integral
$\langle 3d 3d | \hat{u} | 3d 4d \rangle$, corresponding to $F^0$.\cite{comment4}
Results of these calculations are shown
in Fig.~\ref{fig.relaxation}.
The region of $3d$$\rightarrow$$4d$ transitions strongly overlaps with the
``plasmon peak'' of ${\rm Im} W(\omega)$ (Fig.~\ref{fig.WandSb}).
Therefore, in the GW calculations, these two effects
are strongly mixed.\cite{FerdiGunnarsson,Springer,Kotani,Ferdi}
The $\omega$-dependence of $\Delta W$ will also
contribute to the renormalization
of the low-energy part of spectrum. In GW, this contribution can be estimated
using Eq.~(\ref{eqn:dSigmamu}), which yields
$\left. \partial \Sigma /\partial \omega \right|_{\omega=\mu}$$\sim$$0.06$.
This contribution is small
and can be neglected.

\section{\label{sec:GWversusCLDA}GW versus CLDA}

  What is missing in the \textit{ab initio} GW method, and what is the relation
between GW and CLDA?
Let us consider for simplicity the static case, where $\delta \hat{V}_{\rm ext}$
does not depend on time (the generalization to the time-dependent case is rather
straightforward).

  Eventually, both methods are designed to
treat the response $\delta\rho({\bf r})$ of the charge
density (\ref{eqn:KS})
to the change of the external potential $\delta \hat{V}_{\rm ext}$,
which can be calculated
in
the first order of
the regular perturbation theory.
Then,
$\delta \hat{V}_{\rm ext}$ will affect both
eigenvalues and
eigenfunctions of the KS equations (\ref{eqn:KS}). The corresponding corrections are given by
the matrix elements $\langle \psi_i | \delta \hat{V}_{\rm ext} | \psi_j \rangle$
with $i$$=$$j$ and $i$$\neq$$j$, respectively.
If two (or more) eigenvalues are located near the Fermi level, their shift can lead to
the repopulation effects when some levels become occupied at the expense of
the other ones. This is a direct consequence of the conservation of the total
number of particles, which affects the occupation numbers. Therefore,
very generally, the total response $\delta\rho({\bf r})$ in metals will
consist of two parts,
$\delta \rho({\bf r})$$=$$\delta_1 \rho({\bf r})$$+$$\delta_2 \rho({\bf r})$,
describing the change of the occupation numbers,
$\delta_1 \rho({\bf r})$$=$$\sum_i \delta f_i |\psi_i ({\bf r})|^2$,
and the relaxation of the
wavefunction,
$\delta_2 \rho({\bf r})$$=$$\sum_i f_i \delta |\psi_i ({\bf r})|^2$, respectively.
Then, the
polarization function $P$, defines as
\begin{equation}
\delta \rho({\bf r}) = \int d {\bf r}' P({\bf r},{\bf r}',0) \delta V_{\rm ext} ({\bf r}'),
\label{eqn:deltarhoRPA}
\end{equation}
will also
consist of two parts, $P_1$ and $P_2$, which yield
$\delta_1 \rho$ and $\delta_2 \rho$ after acting on $\delta V_{\rm ext}$.
Then, it is easy to verify by considering the perturbation-theory
expansion for $\{ \psi_i \}$
with respect to $\delta V_{\rm ext}$ that the GW approximation corresponds to the choice $P_1$$=$$0$ and
$P_2$$=$$P_{\rm GW}$.
It yields
$\delta_2\rho({\bf r})$, which further induces the new change of the
Coulomb (Hartree) potential
$\delta_2 V_{\rm H}({\bf r})$$=$$e^2 \int d {\bf r}'
\delta_2 \rho({\bf r}')/|{\bf r}$$-$${\bf r}'|$.
By solving this problem self-consistently and taking the functional derivative
with respect to $\delta_2 \rho$ one obtains the GW expression (\ref{eqn:Dyson})
for the screened Coulomb interaction $\hat{W}_{\rm GW}(0)$.
Therefore, it is clear that the \textit{ab initio} GW method takes into account only one part of the
total response $\delta \rho$, describing the relaxation of the wavefunction with the
fixed occupation numbers. Another contribution, corresponding to the change of the
occupation numbers (or the charge redistribution near the Fermi level)
is totally missing.

  This result can be paraphrased in a different way, which clearly illustrates its
connection with the definition of orthogonal subspaces, $L$ and $I$, discussed in the
introduction, and the partitioning of the polarization function $P$ (Sec.~\ref{sec:GWg}),
which is used in the definition of the Hubbard model.\cite{Kotani,Ferdi}
First, recall that according to the main idea of the
LDA$+$$U$ method (see postulates 1-3 of the Introduction part), $\delta \hat{V}_{\rm ext}$
should be a projector-type operator acting in the subspace of the $L$ states.
Then, the result of the action of the polarization function
$P_{\rm GW}$$\equiv$$P_2$, given by
Eq.~(\ref{eqn:polarization}),
onto this
$\delta \hat{V}_{\rm ext}$
will belong to the same $L$ space. Therefore, the projection $\delta \hat{V}_{\rm ext}$
will generate only that part of the polarization function, which is associated with
the transitions between localized states
($\hat{P}_{LL}$ in Sec.~\ref{sec:GWg}). Meanwhile, this polarization effect
should be
excluded from
the final definition of the parameter $U$ in the Hubbard model to avoid the
double counting.\cite{Kotani,Ferdi}
However, if $\hat{P}_{LL}$ is excluded, there will be nothing left in the
polarization
function (\ref{eqn:polarization}) that can interact with
$\delta \hat{V}_{\rm ext}$ and screen the change of the electron density in the
$L$-subspace.
Therefore, the GW scheme should correspond
to the bare Coulomb interaction, that is totally consistent
with the analysis presented in Sec.~\ref{sec:GWNi}.

\subsection{\label{sec:difficulties}Basic Difficulties for Transition Metals}

  There is certain ambiguity in the construction of the Hubbard model for
the transition metals, which is related with the fact that their LDA electronic
structure cannot be described in terms of fully separated $L$- and $I$-states without additional
approximations. In this section we briefly review two such approximations, which will
explain the difference of our point of view on the screening of Coulomb
interactions in the transition metals from the one proposed in
Refs.~\onlinecite{Kotani} and \onlinecite{Ferdi}.

\begin{figure}[h!]
\begin{center}
\resizebox{8.0cm}{!}{\includegraphics{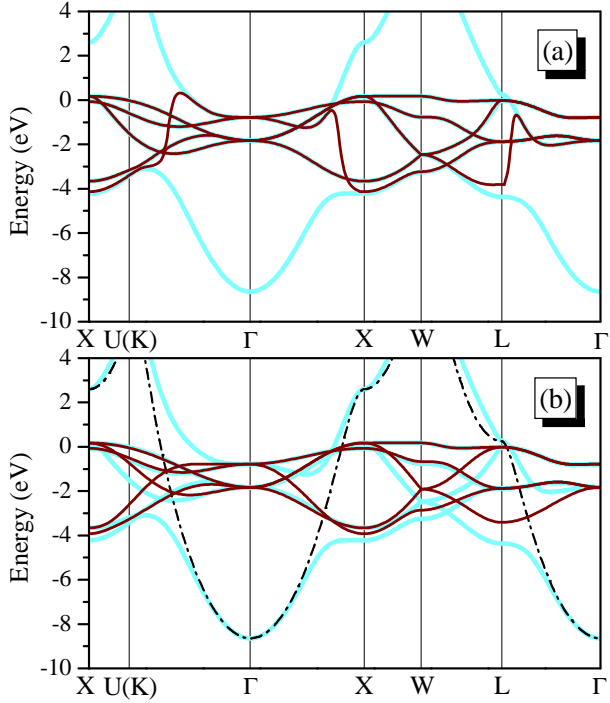}}
\end{center}
\caption{\label{fig.bands}
Two approximate views on the electronic structure of (paramagnetic) fcc Ni
underlying different schemes of calculation of the screened Coulomb interaction.
The original LDA bands are shown by light color.
The GW calculations are based on the model `a', which implies that \textit{all}
electronic structure near the Fermi level (located at zero) can be described in terms
of only \textit{five} pseudo-Wannier orbitals of predominantly $3d$-character,
simulating the $L$-states. The dark bands show an example of such electronic
structure obtained after elimination of $4sp$-states through the downfolding
procedure.\protect\cite{PRB04} The remaining electronic states are the $I$-states,
which are orthogonal
to the pseudo-Wannier orbitals and allowed to screen the Coulomb interactions
in these bands. The screening is treated in RPA.
The model `b', which is used in constraint-LDA calculations, is obtained after
neglecting the hybridization between $3d$- and $4sp$-states (the so-called
canonical-bands approximation~\protect\onlinecite{LMTO}). It consists of the
$3d$ band (representing the $L$-states and shown by dark color), embedded into the free-electron-like
$4sp$-band (representing the $I$-states and shown by dash-dotted line).
The coexistence of two different groups of states near the Fermi level gives rise
to the charge redistribution, which contribute to the screening of Coulomb interactions in
the $3d$ band.
}
\end{figure}

  The GW approach employed in Refs.~\onlinecite{Kotani} and \onlinecite{Ferdi}
implies
that \textit{all} electronic structure near the Fermi level can be described in
terms of \textit{only five} pseudo-Wannier orbitals of predominantly $3d$-character,
which serve as the $L$-states in the considered model.
Generally, such $L$-states are not the same as the LMTO basis functions and
take into account the effects of hybridization
between $3d$ and $4sp$ states.
An example of such electronic
structure, obtained after elimination of the $4sp$-states near the Fermi level through the
downfolding procedure,\cite{PRB04} is shown in Fig.~\ref{fig.bands}.
Other possibilities of defining these pseudo-Wannier functions, which have been
actually used in Refs.~\onlinecite{Kotani} and \onlinecite{Ferdi}, are summarized
in Ref.~\onlinecite{comment3}.
Then, the remaining electronic states, which are orthogonal to these pseudo-Wannier
orbitals, represent the $I$-states. By the construction, the $I$-states are expected
to be far from the Fermi level. This may justify the use of the GW approximation for
the screening of Coulomb interactions in the $3d$-electron-like bands, formed by the
pseudo-Wannier orbitals near the Fermi level, by the remote $I$-states.
The parameters of Coulomb interactions, constructed in such a way,
correspond to the original Herring definition (\ref{eqn:HerringU1}) in the basis
of pseudo-Wannier orbitals. Formally, it should also include the charge redistribution
effects near the Fermi level. However, in this case the charge redistribution
goes between pseudo-Wannier orbitals of the same ($L$) type, which
constitutes the basis of the Hubbard model. Therefore, the effects of the charge
redistribution can be taken into account by including the intra- as well as inter-site Coulomb
interactions in the Hubbard Hamiltonian. The latter can be evaluated in the
GW approach, provided that the relaxation effect are not very sensitive to whether
the excited electron is placed on another
$L$-orbital of the same system, or completely removed from it, like in the GW method.

  The model employed in CLDA calculations is obtained after neglecting
the hybridization between $3d$- and $4sp$-states (the so-called canonical-bands
approximation--Ref.~\onlinecite{LMTO}). It consists of the pure $3d$-band, located
near the Fermi level and representing the $L$-states of the model, which is
embedded into the free-electron-like $4sp$-band, representing the $I$-states.\cite{comment1}
Formally, these bands are decoupled and the free-electron-line $4sp$-band
can be eliminated from the basis in the process of construction of the
Hubbard Hamiltonian. However, in this case the definition of the screened Coulomb
interaction in the $3d$ band should take into account the processes corresponding to
redistribution of electrons between $3d$- and $4sp$-band at the low-energy cost,
which is traced back to Herring's scenario of screening in the transition metals,\cite{Herring}
and which is missing in the GW method.

  However, we would like to emphasize again that both considered models are
\textit{approximations} to the real electronic structure of fcc Ni.
Even in the first case (model `a' in Fig.~\ref{fig.bands}), the free-electron-like
$4sp$-band lies near the Fermi level (especially around L-point of the
Brillouin zone). Therefore, the charge redistribution effects are expected to play some
role even in the basis of Wannier orbitals.
On the other hand, because of strong hybridization between $3d$- and $4sp$-states in the
transition metals, there is a substantial difference of electronic structure used in
CLDA calculations (model `b' in Fig.~\ref{fig.bands})
from the real LDA electronic structure of fcc Ni.
Strictly speaking, all partial contributions to the screening of Coulomb interactions,
which we will consider in the next section, will be evaluated for this particular model of the
electronic structure. The values of these parameters can be revised
to a certain extent after taking into account the hybridization between
$3d$- and $4sp$-states. For example, with the better choice of the Wannier basis
for the five $3d$-electron-line bands
in the model `b' one could possibly incorporate the main effects of the model `a'
and merge these two approaches.

\section{\label{sec:CLDATM}CLDA for transition metals}

  How important are the
relaxation of the wavefunctions and the change of the occupation numbers in the definition
of the Coulomb interaction $U$?
For the transition metals,
both contributions can be easily evaluated in
CLDA. For these purposes it is convenient to use
the Hellman-Feinman theorem, which relates the static $U$ with the expectation value of the KS
potential $V_{\rm KS}$$=$$V_{\rm H}$$+$$V_{\rm XC}$:~\cite{PRB94.2}
$$
U=\langle 3d | \frac{\partial V_{\rm KS}}{\partial n_{3d}} | 3d \rangle.
$$
Then, the exchange-correlation (XC) part is small.
$\delta V_{\rm H}$ can be expressed through $\delta\rho$. Hence, the CLDA
scheme provides the self-consistent solution for $\delta\rho$ associated with the
change of the number of $3d$ electrons, $\delta n_{3d}$. The latter is controlled by
$\delta V_{\rm ext}$. Therefore, the procedure is
totally equivalent to the calculation
of the polarization function $P$ and the screened Coulomb interaction for
$\omega$$=$$0$.

\subsection{\label{sec:conventions}Conventions}

  We use rather standard set up for the CLDA calculations.
Namely, the $3d$ band of Ni should be well separated
from the rest of the spectrum (otherwise, the LDA$+$$U$ strategy
discussed in the Introduction does not apply).
For fcc Ni this is not the case.
However, this property can be enforced by using
the canonical bands approximation in the LMTO method.\cite{LMTO} We employ
even cruder approximation and replace the $3d$ band by the atomic $3d$ levels embedded
into the $4sp$ band (in the other words, we switch off the hybridization
between $3d$ orbitals located at different atomic sites as well as
the $3d$ and $4sp$ states).\cite{AnisimovGunnarsson}
Then, each $3d$ orbital can be assigned to a single atomic site.
By changing the number of $3d$-electrons at different atomic sites $\{ {\bf R} \}$
in supercell calculations,
one can mimic the ${\bf q}$-dependence of the external potential (\ref{eqn:dVext}).
Other atomic population (of the $4sp$ states) are allowed to relax self-consistently
onto each change of the number of $3d$ electrons.
Hence, the contribution of the charge-transfer excitation (\ref{eqn:HerringU2})
to the screening of $U$ is unambiguously defined by the form of the external
potential and details of the electronic structure of the $4sp$ states.
Some aspects of treating the $3d$ states beyond the atomic approximation will
be considered in Sec.~\ref{sec:OQ}.

  The LMTO method is supplemented with an additional spherical
approximation for $V_{\rm KS}({\bf r})$ inside atomic spheres, which bars
small exchange interactions between $3d$ and $4sp$ electrons from the screening of $U$.
By paraphrasing this statement in terms of the polarization function in the GW method,
the spherical approximation for $V_{\rm KS}({\bf r})$ in the CLDA
calculations is equivalent
to retaining in $P_{\rm GW}$ only
those contributions which are associated with
transitions between states with
the same angular momentum (e.g., $3d$$\rightarrow$$4d$, etc.).

\subsection{\label{sec:Gamma}Screened Coulomb Interaction in the $\Gamma$-point}

  First, we evaluate the pure effect associated with the change of the occupation numbers,
without relaxation of the wavefunctions.
This mechanism is directly related with the conservation of the total number of particles,
and simply means that the excess
(or deficiency) of the $3d$ electrons
for ${\bf q}$$=$$0$
is always compensated by the
$4sp$ electrons, which participate in the screening of $3d$ interactions.
The corresponding contribution to the screening of $F^0$ is given by:\cite{PRB94.2}
$$
\Delta^{(1)} F^0 = \sum_{i \neq 3d} \frac{\delta f_i}{\delta n_{3d}}
\langle 3d i|\hat{u}|3d i \rangle_{\rm av}.
$$

  In transition metals, $\Delta^{(1)}U$ is very large and takes into account more than
70\% of screening of the bare Coulomb interaction $F^0$ (Table~\ref{tab:U}). This contribution is
missing in the GW method. The second largest effect ($\sim$25\% of the total screening)
is caused by
relaxation of the $3d$ orbitals
onto the change of the Hartree potential associated with the change of these
occupation numbers
($\Delta^{(2)} U$ in Table~\ref{tab:U}).
The remaining part of the screening ($\sim$5\%) comes from the relaxation of other
orbitals (including the core ones) and the change of the XC potential.
In principle, the relaxation effects should be taken into account by the GW calculations.
However, this procedure strongly
depends on the way how it is implemented.
\begin{table}[h!]
\caption{Partial contributions to the screening of the $3d$ interactions
in the $\Gamma$-point
extracted from constraint-LDA calculations (in eV): (1) bare Coulomb
integral $F^0$, (2) the screening of $F^0$ by the $4sp$ electrons
associated with the
change of occupation numbers,
without relaxation of the wavefunctions ($\Delta^{(1)} F^0$), (3) the
additional screening of $F^0$ associated with relaxation of the $3d$ orbitals
($\Delta^{(2)} F^0$), and (4)
the total value of $U$ obtained in CLDA calculations.} \label{tab:U}
\begin{ruledtabular}
\begin{tabular}{ccccc}
compound & $F^0$ & $\Delta^{(1)} F^0$ & $\Delta^{(2)} F^0$ & $U$ \\
\hline
bcc Fe   & 22.2  & -13.6                 & -3.5            & 4.5            \\
fcc Ni   & 24.9  & -14.2                 & -5.2            & 5.0            \\
\end{tabular}
\end{ruledtabular}
\end{table}
For example,
the CLDA approach is based on a direct solution of KS equations supplemented with
a \textit{flexible} atomic basis set,
like
in the LMTO method.\cite{LMTO}
Then, the change of
$F^0$ caused by relaxation of the $3d$
orbitals can be easily evaluated as\cite{PRB94.2}
$$
\Delta^{(2)} F^0=\frac{n_{3d}}{2} \frac{\partial F^0}{\partial n_{3d}}.
$$
Since $n_{3d}$ is large in the fcc Ni, this contribution is also large.
The situation can be different in the GW scheme, based on the perturbation
theory expansion, which requires a large basis set.\cite{HybertsenLouie}
For example, in order to describe properly the same relaxation of
the $3d$ wavefunctions, the polarization $P_{\rm GW}$ should explicitly
include the excitation from the occupied $3d$ to the unoccupied
$4d$ (and probably higher) states.\cite{FerdiGunnarsson}

\subsection{\label{sec:q}{\bf q}-dependence of Coulomb $U$}

  Since the change of the number of $3d$ electrons in transition metals is \text{not}
totally screened by the $4sp$ electrons at the same atomic site,\cite{AnisimovGunnarsson}
it is reasonable to expect an appreciable ${\bf q}$-dependence of the effective $U$.
Results of CLDA calculations for the high-symmetry points of the Brillouin zone
are summarized in Table~\ref{tab:Uq}. The effective $U$ appears to be small in the
$\Gamma$-point due to the perfect screening by the $4sp$ electrons.
At the Brillouin zone boundary this channel of
screening is strongly suppressed
that is reflected in the larger values of the Coulomb $U$.
The screening by intersite Coulomb interactions, which takes place in the ${\rm X}$-point
of the BZ,
is substantially weaker and cannot fully compensate the lack of the $4sp$-screening.
In the ${\rm L}$-point of the BZ for the fcc lattice,
the modulation of the $3d$-electron density in the CLDA calculations is such that
the number of nearest neighbors with
excessive and deficient number of $3d$ electrons is the same. Therefore, the contributions of
intersite Coulomb interactions to the screening are cancelled out, resulting in the
largest value of the effective $U$ in this point of the BZ.
\begin{table}[h!]
\caption{Coulomb interaction $U$ (in eV) for fcc Ni
in three different points
of the Brillouin zone: $\Gamma$$=$$(0,0,0)$, ${\rm X}$$=$$(2\pi,0,0)$, and
${\rm L}$$=$$(\pi,\pi,\pi)$ (in units of $1/a$, where $a$ is the
cubic lattice parameter).} \label{tab:Uq}
\begin{ruledtabular}
\begin{tabular}{ccc}
$\Gamma$ & ${\rm X}$ & ${\rm L}$ \\
\hline
 5.0                 & 6.8            & 7.3            \\
\end{tabular}
\end{ruledtabular}
\end{table}

\section{\label{sec:GWCLDA}{GW starting with CLDA}}

  In this section we discuss some relevance of parameters of effective Coulomb interactions
extracted from CLDA for the analysis of electronic structure and properties of fcc Ni. We
consider the ``renormalized GW approach'', in which, instead of bare Coulomb interactions,
we use parameters extracted from CLDA. The main difference is that the latter incorporates the
screening by the $4sp$-electrons, including the effects of charge redistribution beyond
the GW approximation. This strategy can be well justified within RPA, because
it allows to partition the polarization function and treat the screening effects in two
steps:\\
(1) We take into account the screening by ``non-$3d$'' electrons using CLDA. This yields
the new (``renormalized'') matrix of screened Coulomb interactions $\hat{\bar{u}}_{LL}$
between the $3d$ electrons.\cite{comment2}
As it was discussed
in Sec.~\ref{sec:q}, the obtained interaction $\hat{\bar{u}}_{LL}$ is
${\bf q}$-dependent, and this dependence is fully taken into account in our calculations.\\
(2) We evaluate the screening caused by $3d$$\rightarrow$$3d$ transitions in the
polarization function (\ref{eqn:polarization}) using Eq.~(\ref{eqn:Dyson}) in which
the matrix of bare Coulomb interactions $\hat{u}_{LL}$ is replaced by $\hat{\bar{u}}_{LL}$.
This yields the new interaction $\hat{\bar{W}}(\omega)$, which is
used in subsequent calculations of the self-energy (\ref{eqn:SigmaGW}).
It is reasonable to expect that the main
$\omega$-dependence of $\hat{\bar{W}}$ will come
from the $3d$$\rightarrow$$3d$ transitions
(see closing
arguments in Sec.~\ref{sec:GWNi}), which are taken into account
in the second step. The screening by ``non-$3d$'' states can be treated as
static.

  Results of these calculations are shown in Fig.~\ref{fig.WandSb}.
\begin{figure}[h!]
\begin{center}
\resizebox{12.0cm}{!}{\includegraphics{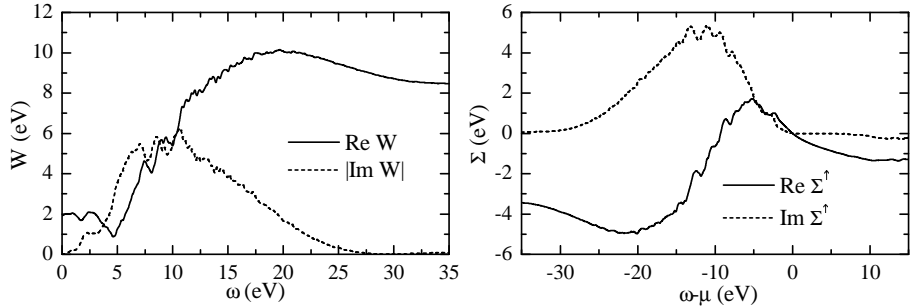}}
\end{center}
\caption{\label{fig.WandSc}
The same as Fig.~\protect\ref{fig.WandSb} but with the parameters of
Coulomb interactions extracted from CLDA.
}
\end{figure}
The main effect of the $4sp$-screening, beyond the standard GW approach,
is the change of the energy scale, which is now controlled by the ${\bf q}$-dependent
Coulomb interaction $U$, being of the order of 5.0-7.3 eV.
It change the asymptotic behavior ${\rm Re} \bar{W}(\infty)$ as well as
the position and the intensity of the ``plasmon peak'' of
${\rm Im} \bar{W}(\omega)$, which is shifted to the lower-energies region
and becomes substantially broader in comparison with the case of bare
Coulomb interactions considered in Sec.~\ref{sec:GWNi}.
On the other hand, the static limit ${\rm Re}\bar{W}$$\simeq$$1.9$ eV
is practically not affected by details of the $4sp$-screening, due to the
strong-coupling regime realized in the low-$\omega$ region. The ${\rm Re} \bar{W}$
exhibits a strong $\omega$-dependence at around 7 eV, which is related with
the position of the plasmon peak of ${\rm Im} \bar{W}(\omega)$.
All these features are well reflected in the behavior of $\Sigma(\omega)$.

  The main effect of the $4sp$-screening onto the spectral function in RPA
consists in somewhat milder reduction of the bandwidth, which is also
related with the spectral weight transfer (Fig.~\ref{fig.DOS}):
the new renormalization factor is $Z$$\sim$$0.7$ against $Z$$\sim$$0.5$
obtained with bare Coulomb interactions. However, the exchange splitting
does not change and the 6 eV satellite structure does not emerge.

\section{\label{sec:OQ}Summary and Remaining Questions}

  We have considered several mechanisms of screening of the bare Coulomb
interactions between $3d$ electrons in transition metals.
We have also discussed different methods of calculations of the screened
Coulomb interactions. Our main results can be summarized as follows.\\
(1) The processes which mainly contribute to the screening of Coulomb interactions
between $3d$ electrons
are essentially local, meaning
that the on-site Coulomb interactions are most efficiently screened by the $3d$ and
$4sp$ electrons located at the same
site.\cite{Gunnarsson,AnisimovGunnarsson,PRB96}
The most efficient mechanism of screening is basically the self-screening
by the same $3d$ electrons, evaluated in some appropriate atomic-like
basis set, like that of the LMTO method employed in the present work.
The $\omega$-dependence
of the effective Coulomb interaction $U$ also originates mainly from the
self-screening.\\
(2) We have clarified a fundamental difference between constraint-LDA and GW
methods in calculating the effective Coulomb interaction $U$. The GW approximation
does not take into account a screening of the on-site Coulomb
interactions by the itinerant $4sp$ electrons, taking place via
redistribution of electrons between $3d$ and $4sp$ bands.

  In a number of cases, the GW approach may be justified by using Wannier basis
functions, representing the bands near the Fermi level. If these bands are
well isolated from the other bands, the redistribution of electrons between
Wannier orbitals for the bands near the Fermi level and those far from the
Fermi level must be negligible.
Then, the remote bands can participate in the screening of Coulomb interactions
in the ``near-Fermi-level bands'' only via virtual excitations, which can be
treated on the RPA level.

  However, in the case of Ni, such separation of bands is not complete, and it is
essential to consider additional mechanisms of screening beyond the GW approximation.
In the present work, the $4sp$-screening is automatically taken into account
in the CLDA approach, which is complementary to the GW method.
Due to the strong-coupling regime realized in RPA equations for the screened
Coulomb interaction, the static limit appears to be insensitive to the details
of the $4sp$-screening. However, from the viewpoint of the
present approach, the $4sp$-screening becomes increasingly important
at finite $\omega$ and controls both the asymptotic behavior and the
position of the plasmon peak of the screened Coulomb interaction in RPA.
The latter effect can be especially important as it predetermines the position
of the satellite structure.

  Finally, we would like to make several comments about implication of the parameters
of screened Coulomb interaction obtained in our work for the description of electronic
structure and properties of transition metals. We will also discuss some future directions
and make a comparison with already existing works. \\
(1) Our results clearly show that RPA is not an adequate approximation for the
electronic structure of fcc Ni. Even after taking into account the additional
screening of the $3d$-$3d$ interactions by the itinerant $4sp$ electrons,
beyond the GW approximation, and the ${\bf q}$-dependence of the effective $U$,
we obtain only a partial agreement with the experimental data. Namely, only the
bandwidth is corrected in this ``renormalized GW approach'', in a better agreement
with the experimental data. However, there is only a tiny change of the spectral
weight around 6 eV (Fig.~\ref{fig.DOS}), i.e. in the region where the satellite
structure is expected experimentally. Even assuming that our parameters of Coulomb
interactions may be still overestimated (due to the reasons which will be discussed
below), and the satellite peak can emerge for some smaller values of $U$,\cite{Ferdi}
one can hardly expect the strong spin-dependence of this satellite structure
as well as the reduction of the exchange splitting, which are clearly seen in the
experiment,\cite{Kakizaki} on the level of RPA calculations.
Therefore, it is essential to go beyond. \\
(2) Even beyond LDA, do the parameters of screened Coulomb interaction
$U$$\sim$5.0-7.3 eV, obtained in the atomic approximation, provide a coherent description
for the electronic structure and properties of fcc Ni?
Probably, this is still an open question because so far not all of the possibilities
in this direction have been fully investigated. One new aspect suggested by
our calculations is the ${\bf q}$-dependence of the effective $U$.
On the other hand, all previous calculations suggest that the Coulomb interaction
of the order of
5.0-7.3 eV is probably too large. For example, the value of $U$, which
provide a coherent description for a number of electronic and magnetic
properties of fcc Ni on the level of DMFT calculations is about 3 eV,\cite{LichtPRL01}
which is well consistent with the previous estimates based on the $t$-matrix
approach.\cite{Liebsch}
Therefore, it is reasonable to ask if
there is an additional mechanism, which further reduces the effective $U$ from
5.0-7.3 eV till 3.0 eV? One possibility lies in the atomic
approximation which neglects the hybridization effects
between $3d$ and $4sp$ states, and which is rather crude
approximation for the transition metals.\cite{comment1}
The hybridization will generally mix the states of the $3d$ and $4sp$ character, and therefore
will affect the form of the Wannier orbitals constructed from the atomic wavefunctions.
Since the $3d$, $4s$, and $4p$ states belong to different representations of
point group of the cubic symmetry, they cannot mix at the same site.
However, the $4s$ (or $4p$) orbital can have tails of the $3d$ character at the
neighboring sites (and vice versa). These tails will additionally screen the Coulomb
interactions between the (nominally) $3d$ electrons. The screening is expected to be very
efficient because it operates between orbitals of the same ($3d$) type.
It should explain further reduction of the static $U$ obtained in the atomic approximation.
Another feature of this screening is the $\omega$-dependence of the effective $U$,
which comes from the $3d$$\rightarrow$$3d$ transitions in the polarization function
(namely between tails of the $4sp$-orbitals and the heads of the wavefunctions of
the $3d$ character). In RPA, this $\omega$-dependence is directly related with the
static limit of screening via the Kramers-Kronig transformation.\cite{FerdiGunnarsson}
We believe that the screening by the tails of the Wannier functions was the main
physical mechanism underlying the calculations of effective Coulomb interaction
in Refs.~\onlinecite{Kotani,Ferdi}, in the framework of \textit{ab initio} GW method,
although this idea has not been clearly spelled out before.
The effect of charge redistribution between different states located near the
Fermi level, which is not taken into account in the GW approximation, is also
expected to be smaller with the proper choice of the Wannier orbitals.

  Another problem is that the $3d$ and $4sp$ bands are strongly mixed in the
case of pure transition metals. Therefore, the construction of the separate
Wannier functions of the ``$3d$'' and ``non-$3d$'' type will always suffer
from some ambiguities.\cite{comment3}
In this sense, the transition-metal oxides, whose physical properties are mainly
predetermined by the behavior of
a limited number of $3d$ bands, located near the Fermi level and well
separated from the rest of the spectrum, are much more interesting systems for
the exploration of the idea of screening of Coulomb
interactions, formulated on the Wannier basis.
For example, based on the above argument, one can expect a very efficient screening
of Coulomb interactions in the $3d$ band by the Wannier states constructed from the
oxygen $2p$ orbitals, which have appreciable tails of the $3d$ character at the transition-metal sites.
The first attempt to consider this screening
have been undertaken in Ref.~\onlinecite{PRB96}, on the basis of
constraint-LDA method. Similar scheme can be formulated within RPA, which takes
into account the $\omega$-dependence of the screened Coulomb
interaction $U$. This work is currently in progress.


\begin{thebibliography}{10}
\bibitem{TDDFT}
An alternative is to calculate the response to the time-dependent (TD)
external potential using TDDFT.\cite{OnidaReiningRubio}
\bibitem{comment3}
The scheme proposed by Kotani (Ref.~\onlinecite{Kotani})
was based on a special construction of the LMTO basis set.\cite{LMTO}
In the process of calculations of the screened Coulomb interaction
$\hat{W}_r$, he excluded only the ``heads'' of the LMTO
$3d$-basis functions, $\phi_{3d}$ (solutions of KS equations inside
atomic spheres for a given energy $E_\nu$), whereas the ``tails''
of the wavefunctions of the same $3d$ character, given by the energy
derivatives $\dot{\phi}_{3d}$ of $\phi_{3d}$, were allowed to screen
$\hat{W}_r$. The idea was probably to go beyond the atomic approximation
for the $3d$ states and to mimic the behavior of the Wannier functions.
However, this procedure depends on the choice of $\{ E_\nu \}$
and the LMTO representation.\cite{LMTO}
For example in the (most suitable for the construction of the Hubbard
model) orthogonal representation, both $\phi_{3d}$ and $\dot{\phi}_{3d}$
can contribute to the ``heads'' of the basis functions as well as
their ``tails''. On the other hand, Aryasetiawan and co-workers in Ref.~\onlinecite{Ferdi}
used the band picture and considered the screening of the ``$3d$'' band
by the ``non-$3d$'' (mainly $4sp$) ones. In reality, however, the
$4sp$ band can carry a considerable weight of atomic $3d$ states, due to the
hybridization. Moreover, the $3d$ and $4sp$ bands in the fcc Ni are well
separated only along the $X$-$\Gamma$-$L$ direction of the BZ. Along other
directions, there is a substantial band-crossing (see Fig.~\ref{fig.bands})
and this procedure is not well
defined.
\bibitem{comment4}
The $4d$ and $5d$ states were treated in the canonical bands approximation~\cite{LMTO}.
Since the energy distance between the $3d$ and $4d$ (and higher) bands is much larger than
the $3d$ bandwidth, one can neglect the dispersion of the $3d$ bands and relate
$P_{\rm GW}(\omega,3d$$\rightarrow$$nd)$ with the local density of $nd$ states.
\bibitem{comment1}
In fact the considered example of the electronic structure where there is a narrow band,
formed by some patricular group of ($L$) states, embedded into the free-electron-like band
is rather generic and equally apply to transition, rear earth and actinide metals.
The main difference between these systems is the strength of hybridization between
$L$- and $I$-states (which is the strongest in the case of transition metals), as well as
in the system of $L$-states.
\bibitem{comment2}
In addition to the parameter $U$, which has a meaning of screened Slater's integral $F^0$,
CLDA allows to easely calculate the intraatomic exchange coupling $J$.\cite{AZA}
The other two parameters, $F^2$ and $F^4$, which specify the matrix $\hat{\bar{u}}_{LL}$
of screened Coulomb interactions between the $3d$ electrons, can be estimated from $J$
using the ratio $F^4/F^2$$\simeq$$0.63$ corresponding to the
atomic limit.\cite{PRB94}
\end{thebibliography}
\end{document}